\documentstyle[12pt]{article}

\def\hybrid{\topmargin -20pt    \oddsidemargin 0pt
        \headheight 0pt \headsep 0pt
        \textwidth 6.25in       
        \textheight 9.5in       
        \marginparwidth .875in
        \parskip 5pt plus 1pt   \jot = 1.5ex}

\hybrid

\def\baselinestretch{1.2}

\catcode`\@=11

\def\marginnote#1{}
\newcount\hour
\newcount\minute
\newtoks\amorpm
\hour=\time\divide\hour by60
\minute=\time{\multiply\hour by60 \global\advance\minute by-\hour}
\edef\standardtime{{\ifnum\hour<12 \global\amorpm={am}%
        \else\global\amorpm={pm}\advance\hour by-12 \fi
        \ifnum\hour=0 \hour=12 \fi
        \number\hour:\ifnum\minute<10 0\fi\number\minute\the\amorpm}}
\edef\militarytime{\number\hour:\ifnum\minute<10 0\fi\number\minute}

\def\draftlabel#1{{\@bsphack\if@filesw {\let\thepage\relax
   \xdef\@gtempa{\write\@auxout{\string
      \newlabel{#1}{{\@currentlabel}{\thepage}}}}}\@gtempa
   \if@nobreak \ifvmode\nobreak\fi\fi\fi\@esphack}
        \gdef\@eqnlabel{#1}}
\def\@eqnlabel{}
\def\@vacuum{}
\def\draftmarginnote#1{\marginpar{\raggedright\scriptsize\tt#1}}

\def\draft{\oddsidemargin -.5truein
        \def\@oddfoot{\sl preliminary draft \hfil
        \rm\thepage\hfil\sl\today\quad\militarytime}
        \let\@evenfoot\@oddfoot \overfullrule 3pt
        \let\label=\draftlabel
        \let\marginnote=\draftmarginnote
   \def\@eqnnum{(\theequation)\rlap{\kern\marginparsep\tt\@eqnlabel}%
\global\let\@eqnlabel\@vacuum}  }

\def\preprint{\twocolumn\sloppy\flushbottom\parindent 2em
        \leftmargini 2em\leftmarginv .5em\leftmarginvi .5em
        \oddsidemargin -.5in    \evensidemargin -.5in
        \columnsep .4in \footheight 0pt
        \textwidth 10.in        \topmargin  -.4in
        \headheight 12pt \topskip .4in
        \textheight 6.9in \footskip 0pt
        \def\@oddhead{\thepage\hfil\addtocounter{page}{1}\thepage}
        \let\@evenhead\@oddhead \def\@oddfoot{} \def\@evenfoot{} }

\def\numberbysection{\@addtoreset{equation}{section}
        \def\theequation{\thesection.\arabic{equation}}}

\def\underline#1{\relax\ifmmode\@@underline#1\else
        $\@@underline{\hbox{#1}}$\relax\fi}

\def\titlepage{\@restonecolfalse\if@twocolumn\@restonecoltrue\onecolumn
     \else \newpage \fi \thispagestyle{empty}\c@page\z@
        \def\thefootnote{\fnsymbol{footnote}} }

\def\endtitlepage{\if@restonecol\twocolumn \else \newpage \fi
        \def\thefootnote{\arabic{footnote}}
        \setcounter{footnote}{0}}  

\catcode`@=12
\relax

\def\figcap{\section*{Figure Captions\markboth
        {FIGURECAPTIONS}{FIGURECAPTIONS}}\list
        {Figure \arabic{enumi}:\hfill}{\settowidth\labelwidth{Figure
999:}
        \leftmargin\labelwidth
        \advance\leftmargin\labelsep\usecounter{enumi}}}
 \relax
\def\tablecap{\section*{Table Captions\markboth
        {TABLECAPTIONS}{TABLECAPTIONS}}\list
        {Table \arabic{enumi}:\hfill}{\settowidth\labelwidth{Table
999:}
        \leftmargin\labelwidth
        \advance\leftmargin\labelsep\usecounter{enumi}}}
 \relax
\def\reflist{\section*{References\markboth
        {REFLIST}{REFLIST}}\list
        {[\arabic{enumi}]\hfill}{\settowidth\labelwidth{[999]}
        \leftmargin\labelwidth
        \advance\leftmargin\labelsep\usecounter{enumi}}}
 \relax
\makeatletter
\newcounter{pubctr}
\def\publist{\@ifnextchar[{\@publist}{\@@publist}}
\def\@publist[#1]{\list
        {[\arabic{pubctr}]\hfill}{\settowidth\labelwidth{[999]}
        \leftmargin\labelwidth
        \advance\leftmargin\labelsep
        \@nmbrlisttrue\def\@listctr{pubctr}
        \setcounter{pubctr}{#1}\addtocounter{pubctr}{-1}}}
\def\@@publist{\list
        {[\arabic{pubctr}]\hfill}{\settowidth\labelwidth{[999]}
        \leftmargin\labelwidth
        \advance\leftmargin\labelsep
        \@nmbrlisttrue\def\@listctr{pubctr}}}
 \relax
\makeatother
\newskip\humongous \humongous=0pt plus 1000pt minus 1000pt

\newif\ifdtup

\relax

\def\be{\begin{equation}}
\def\ee{\end{equation}}
\def\ba{\begin{eqnarray}}
\def\ea{\end{eqnarray}}

\def\del{\partial}

\def\r{\rho}

\def\G{\Gamma}
\def\d{\delta}
\def\D{\Delta}
\def\e{\epsilon}

\def\P{\Pi}

\def\th{\theta}

\def\m{\mu}
\def\n{\nu}

\def\Om{\Omega}
\def\l{\lambda}
\def\L{\Lambda}
\def\s{\sigma}

\def\cL{{\cal L}}

\def\cg{{\cal{G} }}
\def\cD{{\cal{D} } }

\def\no{\noindent}

\def\IR{\relax{\rm I\kern-.18em R}}

\def \ha {{1\over 2}}

\def \ov {\over}

\def\IR{\relax{\rm I\kern-.18em R}}
\def\inv{^{\raise.15ex\hbox{${\scriptscriptstyle -}$}\kern-.05em 1}}

\def\PL{Poisson--Lie T-duality}

\def\cL{{\cal L}}

\def\tX{{\tilde X}}

\def\tP{{\tilde P}}
\def\tL{{\tilde L}}

\begin{document}

\renewcommand{\theequation}{\arabic{equation}}

\newcommand{\beq}{\begin{equation}}
\newcommand{\eeq}[1]{\label{#1}\end{equation}}
\newcommand{\ber}{\begin{eqnarray}}
\newcommand{\eer}[1]{\label{#1}\end{eqnarray}}
\newcommand{\eqn}[1]{(\ref{#1})}
\begin{titlepage}
\begin{center}

\hfill CERN-TH/97-285\\
\hfill hep-th/9710163\\

\vskip .8in

{\large \bf CANONICAL EQUIVALENCE OF NON-ISOMETRIC $\s$-MODELS}
{\large \bf AND POISSON--LIE T-DUALITY }

\vskip 0.6in

{\bf Konstadinos Sfetsos
\footnote{e-mail address: sfetsos@mail.cern.ch}}\\
\vskip .1in

{\em Theory Division, CERN\\
     CH-1211 Geneva 23, Switzerland}\\

\vskip .2in

\end{center}

\vskip .6in

\begin{center} {\bf ABSTRACT } \end{center}
\begin{quotation}\noindent

\no
We prove that a transformation, conjectured in our previous work,
between phase-space variables in $\s$-models related by 
Poisson--Lie T-duality is indeed a canonical one.
We do so by explicitly demonstrating the invariance of the 
classical Poisson brackets. This is the first example of a class
of $\s$-models with no isometries related by canonical transformations.
In addition we discuss generating functionals of canonical transformations
in generally non-isometric, bosonic and supersymmetric $\s$-models
and derive the complete set of conditions that determine them.
We apply this general formalism to find the generating functional for 
Poisson--Lie T-duality. We also comment on the relevance of this work to
D-brane physics and to quantum aspects of T-duality.

\vskip .2in

\noindent

\end{quotation}
\vskip 2cm
CERN-TH/97-285\\
October 1997\\
\end{titlepage}
\vfill
\eject

\def\baselinestretch{1.2}
\baselineskip 16 pt

\noindent


\section{Introduction}

Among the different dualities in string theory T-duality \cite{BUSCHER} 
remains by far the best understood and, as such, it has influenced
the development of the subject to a great extent. An elegant, as well as
suitable for many applications, formulation of T-duality is as a 
canonical transformation on phase space.
This was realized first for toroidal constant backgrounds \cite{can1}. 
The ultimate
goal was and still is to extend the canonical equivalence for
curved backgrounds, and eventually even to classify 
them.\footnote{Historically, the development of T-duality on curved
backgrounds followed mainly
the path integral approach. Besides the works we will explicitly mention,
certain aspects of Abelian duality can 
be found in \cite{abel} and of non-Abelian duality in \cite{nonabel}. 
For review articles, see \cite{revtd}.}
The first step in this direction was done for the non-Abelian dual of
the $SU(2)$ Principal Chiral Model (PCM)
with respect to the left (or right) action of $SU(2)$ \cite{zacloz} 
and for the duals of general backgrounds with respect to an
Abelian isometry \cite{can2}.
It was further realized that the canonical transformation for the $SU(2)$
PCM generates, with minor modifications, non-Abelian dualities for
PCMs for any group \cite{can3}
and, remarkably, even for the most general $\s$-model with a non-Abelian
group of isometries acting from the left (or the right) only \cite{KSsusynab}.
There is also formulation of non-Abelian duality 
in WZW models as a canonical transformation \cite{KSsusynab}
(for some relevant remarks, see section 4). This has new interesting 
features since, unlike the previous cases, the duality group acts 
vectorially from the left and the right.
The common feature of the works we have mentioned so far
is that in order to implement the T-duality, either from the path integral
point of view or via the canonical approach, a group of isometries was
needed. In \cite{KliSevI} the situation was improved by generalizing the
concept of T-duality in the absence of
isometries when the corresponding backgrounds fulfil certain conditions.
This kind of T-duality, called \PL,
encompasses all Abelian and non-Abelian T-dualities we have
mentioned, except that for WZW models. In \cite{DriDou} it was shown that 
$\s$-models related by \PL\ may be seen as arising 
from a single action, thus hinting towards
their canonical equivalence. However, the 
generating functional was obtained in an abstract form not suitable for 
extracting the explicit canonical map in phase space.
Such a map was conjectured in \cite{PLsfe1} on the basis of symmetry arguments
and correct limit in the case of the usual non-Abelian duality.

In this paper we are 
going to prove the canonical equivalence by explicitly demonstrating the
invariance of the Poisson brackets under the so far conjectured transformation.
Hence, we provide the first example of a class of $\s$-models that are,
although not isometric, canonically equivalent and put \PL\ on an 
equal footing with
Abelian and non-Abelian duality, for which an explicit canonical formulation 
existed, as explained. 
In addition, we will systematically
formulate the generating functional approach to canonical transformations 
for general bosonic and supersymmetric $\s$-models that do not necessarily 
have isometries. We will apply this general formalism to find perturbatively,
in a sense to be explained in the text, the generating functional for
\PL. The initial motivation for the present work, 
besides the disentanglement of the
existence of a canonical transformation on the existence of isometries, 
stems from the fact that certain phenomena
that present a problem from a field theoretical point of view
are simply ``gauge'' artefacts from a string theoretical view point.
This has been illustrated, by means of the interplay between 
world-sheet supersymmetry and Abelian \cite{basfe} as well as
non-Abelian T-duality \cite{KSsusynab}, with the help of the canonical 
formulation; one would like to extend the discussion
to \PL\ (for work and discussion in this direction see \cite{PLsfe1}). 
The revival of open string theory sparked by \cite{Pol}, thanks
to its relevance
in understanding non-perturbative string dualities, brings up the interest 
in T-duality for open strings \cite{GrHo}, in relation to D-branes
\cite{DaLePo}.
Our results in this paper may help to understand backreaction effects
of the D-branes due to a Fischler--Susskind type mechanism \cite{FiSu},
as we explain in section 4.

The organization of this paper is as follows:
In section 2 we prove the canonical transformation for \PL\ by showing
the invariance of the classical Poisson brackets.
In section 3 we discuss generating
functionals in general bosonic and $N=1$ supersymmetric $\s$-models and 
apply for \PL.
In section 4 we conclude with an extensive discussion of possible 
future directions of this work. 
We have also written in appendix A the derivation of 
some non-trivial identities used in 
the proofs of section 2, and in appendix B 
the derivation of a manifestly \PL\ invariant affine algebra.

\section{The canonical transformation}

\subsection{A brief review}

We consider a 2-dimensional non-linear $\s$-model for the $d$
field variables $X^M=(X^\m,Y^i)$,
where $X^\m$, $\m=1,2,\dots  ,\dim(G)$ parametrize an element
$g$ of a group $G$ and the rest are denoted by 
$Y^i$, $i=1,2,\dots , d-\dim(G)$. 
We also introduce representation matrices
$\{T_a\}$, with $a=1,2,\dots, \dim(G)$ and
the components of the left-invariant Maurer--Cartan forms 
$ L^a_\pm = L^a_\m \del_\pm X^\m$, with an inverse denoted by $L^\m_a$.
For notational convenience we will also use $L^i_\pm = \del_\pm Y^i$ 
and $L^A_\pm = (L^a_\pm,L^i_\pm)$.
The light-cone coordinates on the world-sheet are
$\s^\pm =\ha (\tau \pm \s)$. The 
corresponding action has the form
\ba S 
& = & \ha \int L^A_+E^+_{AB} L_-^B
\nonumber \\
&  = &  \ha \int L_+^a E^+_{ab} L_-^b
+ \del_+ Y^i \Phi^+_{ia} L_-^a 
+ L_+^a \Phi^+_{ai} \del_- Y^i + \del_+ Y^i \Phi_{ij} \del_- Y^j~ .
\label{smoac}
\ea
Similarly we consider another $\s$-model for the $d$
field variables $\tilde X^M=(\tilde X^\m,Y^i)$, where
$\tilde X^\m$, $\m=1,2,\dots  ,\dim(G)$ parametrize a different group 
$\tilde G$, whose dimension is, however, equal to that of $G$.
The rest of the variables are the 
same $Y^i$'s used in \eqn{smoac}. Accordingly we introduce a different set 
of representation matrices $\{\tilde T^a\}$, with $a=1,2,\dots, \dim(G)$ and
the corresponding components of the left-invariant Maurer-Cartan forms 
$\tilde L_{\pm a} =\tilde  L_{\m a} \del_\pm \tilde X^\m$ 
(with inverse $\tilde L^{\m a}$)
as well as $\tilde L_{\pm A} =(\tilde L_{\pm a },L^i_\pm)$.
The corresponding action has the form
\ba \tilde S 
& = & \ha \int \tilde L_{+A} \tilde E^{+AB} \tilde L_{-B} 
\nonumber \\ 
& = & \ha \int \tilde L_{+a} \tilde E^{+ab} \tilde L_{-b} + 
\del_+ Y^i \tilde \Phi^+_i{}^a \tilde L_{-a} 
+ \tilde L_{+a} \tilde \Phi^{+a}{}_i \del_- Y^i + 
\del_+ Y^i \tilde \Phi_{ij} \del_- Y^j~ .
\label{tsmoac}
\ea
The matrix $E^+_{AB}$
(for later use we also define $E^-_{AB}=E^+_{BA}$) in \eqn{smoac} 
may depend on all variables $X^\m$ and $Y^i$ and similarly for the 
matrix $\tilde E^{+AB}$ in the action \eqn{tsmoac}.
Hence, we do not require any isometry associated 
with the groups $G$ and $\tilde G$. 

The $\s$-models \eqn{smoac} and \eqn{tsmoac} 
will be dual to each other 
in the sense of Poisson--Lie T-duality \cite{KliSevI}
if the associated algebras $\cg$ and $\tilde \cg$ form a pair 
of maximally isotropic subalgebras into which the Lie algebra $\cD$ of a 
Lie group $D$ known as the Drinfeld double can be decomposed.
This implies that besides the usual commutators 
\be
 [T_a,T_b] = i f_{ab}{}^c T_c ~ , ~~~
 [\tilde T^a,\tilde T^b] = i \tilde f^{ab}{}_c \tilde T^c ~ ,
\label{f1}
\ee 
for the algebras $\cg$ and $\tilde \cg$, there is also the ``mixed''
commutator to consider
\be
[T_a,\tilde T^b ] = 
i \tilde f^{bc}{}_a T_c - i f_{ac}{}^b \tilde T^c ~ .
\label{mixbra}
\ee
There is also a bilinear invariant 
$\langle{\cdot|\cdot \rangle}$ with the various generators obeying 
\be
\langle{T_a|T_b\rangle}= \langle{\tilde T^a|\tilde T^b \rangle}= 0~ ,
~~~~ \langle{T_a|\tilde T^b \rangle} = \d_a{}^b ~ .
\label{bili}
\ee
For more mathematical details one should consult the 
literature (for instance \cite{alemal}).
\no
There remains to relate $E^+_{AB}$ in \eqn{smoac} to $\tilde E^{+AB}$
in \eqn{tsmoac}.
In order to do that it is
convenient to define matrices $a(g)$, $b(g)$ and $\Pi(g)$ as
\be
 g\inv T_a g = a_a{}^b T_b~ ,~~~~ g\inv \tilde T^a g = 
b^{ab} T_b +  (a\inv)_b{}^a \tilde T^b~ ,~~~~ \Pi^{ab}=
b^{ca} a_c{}^b ~ .
\label{abpi}
\ee
Consistency restricts them to obey\footnote{In this
paper explicit display of all the indices in the various symbols 
will be avoided whenever it causes no confusion.} 
\be
a(g\inv) = a\inv(g)~ ,~~~~ b^T(g)= b(g\inv)~ ,~~~~
\Pi^T(g) = - \Pi(g) ~ .  
\label{conss}
\ee 
Similarly we define matrices 
$\tilde a(\tilde g)$, $\tilde b(\tilde g)$ and $\tilde \Pi(\tilde g)$ 
by just replacing the untilded symbols by tilded ones.
Then the various couplings in the $\s$-model actions 
\eqn{smoac}, \eqn{tsmoac} 
are restricted to be of the form \cite{KliSevI,TyuUng}:
\ba 
&& \Phi^\pm = E^\pm (E_0^\pm)\inv F^\pm ~ ,~~~~ 
E^\pm = \left( (E_0^\pm)\inv \pm \Pi\right)\inv ~ ,\nonumber \\ 
&& \Phi = F -  F^+ \Pi E^+ (E^+_0)\inv F^+ ~ ,
\label{coupl}   
\ea
and 
\ba 
&& \tilde \Phi^\pm = \pm \tilde E^\pm F^\pm ~ ,~~~~~
\tilde E^\pm = \left(E_0^\pm \pm \tilde \Pi\right)\inv ~ , \nonumber \\ 
&& \tilde \Phi = F - F^+ \tilde E^+ F^+ ~ ,
 \label{coupldu}
\ea
where  
$F^+_{ia}=F^-_{ai}$, $F^+_{ai}=F^-_{ia}$ and $(E_0^+)_{ab}= (E_0^-)_{ba}$ 
may be at most functions of the variables $Y^i$ only.
The common parametrization of $E^+_{AB}$ and $\tilde E^{+AB}$ in terms 
of $E^+_0$ and $F^\pm_{ai}$ provides the required relation between them.
Various aspects of \PL\ can be found in \cite{KliSevI,DriDou,TyuUng,
KliSevII,kliDbr,DriDou}, generalization to $N=1$ world-sheet supersymmetric
$\s$-models in \cite{PLsfe1,Klisusy}, and some relations to $N=2$ 
superconformal WZW models in \cite{Parkh}.
 
\subsection{The proof}

In \cite{PLsfe1} it was conjectured, on the basis of symmetry arguments, and
of a correct limit in the case of non-Abelian duality, when 
one of the groups $G$ or $\tilde G$ is Abelian,
that the classical canonical transformation relating the $\s$-model
actions \eqn{smoac} and \eqn{tsmoac} is given by
\ba
L_\s & = & (I - \Pi \tilde \Pi) \tilde P - \Pi \tilde L_\s ~ , 
\nonumber \\
P & = & \tilde \Pi \tilde P +  \tilde L_\s ~ ,
\label{cantr}
\ea
with $P_a = L^\m_a P_\m$, where 
$P_\m= \d S/\d\dot X^\m$ is the conjugate to $X^\m$ momentum, $L^a_\s=
L^a_\m \del_\s X^\m$ and similarly for the corresponding tilded symbols.
We may cast \eqn{cantr} into the equivalent forms, 
which are in addition \PL-symmetric,  
\ba
P & = & \tilde \Pi \tilde P +  \tilde L_\s ~ ,
\nonumber \\
\tilde P & = & \Pi P +  L_\s ~ ,
\label{cantr1}
\ea
and
\ba
P & = & ( I -\tilde \Pi \Pi)\inv 
(\tilde L_\s +  \tilde \Pi L_\s  ) ~ , 
\nonumber \\
\tilde P & = & ( I -\Pi \tilde \Pi)\inv 
 ( L_\s +  \Pi \tilde L_\s ) ~ .
\label{cantr2}
\ea
It was further checked \cite{PLsfe1} that \eqn{cantr} is
good enough to prove that the Hamiltonians corresponding to actions
\eqn{smoac} and \eqn{tsmoac} are equal, i.e. $H = \tilde H$. 
This was already a non-trivial check
since the right-hand side of \eqn{cantr} depends on the $\tX^\m$'s
and the $X^\m$'s and it is not a priori obvious that when \eqn{cantr} is used 
in transforming $H$ all $X^\m$-dependence will cancel out, as 
it eventually does so.
To complete the proof that one of \eqn{cantr}--\eqn{cantr2}
is indeed a canonical transformation,
one has to derive it from a 
generating functional $F$ (with ${\del F\ov \del \tau}=0 $) 
or equivalently prove that 
one of \eqn{cantr}--\eqn{cantr2} leaves invariant
the classical Poisson brackets 
of phase-space variables for
\eqn{smoac} and \eqn{tsmoac}. We found it more convenient to follow
the latter approach, which we present now, leaving for later in the 
paper the discussion of the generating functional.
First, let us focus on the action \eqn{smoac}. The usual pair of phase-space 
variables $(X^\m,P_\n)$ satisfies the basic equal-time Poisson 
brackets\footnote{We will not explicitly display the 
world-sheet dependence of the phase-space variables involved in the various 
Poisson brackets. 
It is understood that the first one in the bracket is always 
evaluated at $\s$ and the second one at $\s'$,
whereas the $\tau$ dependence is common.}:
\ba 
&& \{ X^\m, P_\n\}=\d^\m_\n \d(\s-\s') ~ ,
\nonumber \\
&& \{ X^\m, X^\n\}=\{ P_\m, P_\n\}=0~ .
\label{Poi1}
\ea
Using them and the property of the Maurer--Cartan form $\del_{[\m} L^a_{\n]}
= f_{bc}{}^a L^b_\m L^c_\n$, one computes the equal-time Poisson brackets 
for $(L^a_\s,P_b)$
\ba
\{ L^a_\s,P_b\} & = & f_{bc}{}^a L^c_\s \d(\s-\s') 
+ \d_b{}^a \d^\prime(\s-\s') ~ ,
\nonumber \\
\{ P_a, P_b\} & = & f_{ab}{}^c P_c \d(\s-\s') ~ ,
\label{LP} \\
\{ L^a_\s, L^b_{\s'}\} & = & 0 ~ .
\nonumber
\ea
A simple inspection of the equivalent transformations \eqn{cantr}--\eqn{cantr2}
shows that it will be convenient to consider Poisson brackets for $(J^a,P_b)$,
where
\be
J^a= L^a_\s + \Pi^{ab}P_b ~ .
\label{Ja}
\ee
The reason is that \eqn{cantr1} then simply reads $P_a=\tilde J_a$
and $\tilde P^a= J^a$, with both sides of the transformation depending
on variables defined in \eqn{smoac} or \eqn{tsmoac} only. Notice that this
is not the case for \eqn{cantr} and \eqn{cantr2}.
A relatively simple computation shows that\footnote{Our 
symmetrization and antisymmetrization conventions are $(ab)=ab+ba$ and
$[ab]=ab-ba$.}
\be
\{J^a,P_b\}=\left( f_{bc}{}^a J^c + 
(f_{bd}{}^{[a} \Pi^{c]d} + L^\m_b \del_\m \Pi^{ac} )P_c \right) \d(\s-\s') 
+ \d_b{}^a \d'(\s-\s') ~ ,
\label{JP}
\ee
and 
\ba
\{J^a,J^b\} & = & 
- (f_{cd}{}^{[a} \Pi^{b]d} + L^\m_c \del_\m \Pi^{ab}) L^c_\s \d (\s-\s') 
\nonumber \\
&  &  - (\Pi^{db} \Pi^{ec} f_{de}{}^a + \tilde f^{ac}{}_d \Pi^{bd}
+ \Pi^{ad} L^\m_d \del_\m \Pi^{bc} ) P_c \d(\s-\s') ~ .
\label{JJ}
\ea
Using \eqn{id6} in \eqn{JP} and \eqn{id7}, \eqn{id3} in \eqn{JJ}
we find that they simplify to 
\ba
\{J^a,J^b\} & = & \tilde f^{ab}{}_c J^c \d(\s-\s') ~ ,
\nonumber \\
\{P_a,P_b\} & = & f_{ab}{}^c P_c \d(\s-\s')  ~ ,
\label{JPJP} \\
\{J^a,P_b\} & = & ( f_{bc}{}^a J^c - \tilde f^{ac}{}_b P_c ) \d(\s-\s') 
+ \d_b{}^a \d^\prime(\s-\s') ~ ,
\nonumber
\ea
where we have carried over from \eqn{LP} the expression for the Poisson
bracket $\{P_a,P_b\}$.
It is clear that the analogous equal-time Poisson brackets for the pair 
of phase-space variables $(\tilde J_a,\tilde P^b)$ defined for 
the action \eqn{tsmoac} are
\ba 
\{\tilde J_a,\tilde J_b\}& =& f_{ab}{}^c \tilde J_c \d(\s-\s') ~ ,
\nonumber \\
\{\tilde P^a,\tilde P^b\}& = & \tilde f^{ab}{}_c \tilde P^c \d(\s-\s')  ~ ,
\label{tJPJP} \\
\{\tilde J_a,\tilde P^b\}& =& ( \tilde f^{bc}{}_a \tilde J_c 
- f_{ac}{}^b \tilde P^c ) \d(\s-\s') 
+ \d_a{}^b \d^\prime(\s-\s') ~ .
\nonumber
\ea
Under the transformation $P_a=\tilde J_a$,
$J^a=\tilde P^a$, one goes from \eqn{JPJP} to \eqn{tJPJP} and vice versa,
hence proving that \eqn{cantr1} and equivalently \eqn{cantr} and \eqn{cantr2}
are indeed canonical transformations.
We may view \eqn{JPJP}, \eqn{tJPJP} as the infinite-dimensional analogues
of the group theory duality-invariant 
commutation relations \eqn{f1}, \eqn{mixbra}.

In appendix B
we find a \PL\ invariant affine algebra using the manifestly
\PL\ Lorentz-non-invariant action of \cite{DriDou} as it was modified to 
consistently include the inert fields $Y^i$ in \cite{PLsfe1}.

In subsection 3.2 we will extend the classical canonical equivalence to
$N=1$ world-sheet supersymmetric $\s$-models related by \PL\ 
by finding how the world-sheet fermions transform (see also \cite{PLsfe1}).

\section{On generating functionals}

As we have mentioned, it is not necessary to have a generating functional in
order to prove that \eqn{cantr}--\eqn{cantr2} are canonical transformations,
since we have explicitly shown this to be the case
at the Poisson bracket level.
Nevertheless, it will be nice to obtain it, 
as it may provide more insight into 
\PL\ as well as in searching for the quantum counterpart of the 
classical canonical 
transformation. We will be able to derive it perturbatively, in a sense 
to be explained shortly, but in the process we will provide the general 
setting for determining a generating functional for bosonic and supersymmetric
models once a transformation between phase-space variables is given. 
For bosonic models there is some discussion in \cite{oalvII},
but here we will be explicit.

\subsection{The bosonic case}

Let us consider two $\s$-models with actions
\be
S = \ha \int Q^+_{MN} \del_+ X^M \del_- X^N~ , ~~~ 
Q^\pm_{MN}=G_{MN}\pm B_{MN}~ , ~~~ X^M=(X^\m,Y^i) ~ ,
\label{SQ}
\ee
and
\be
\tilde S = \ha \int \tilde Q^+_{MN} \del_+ \tilde X^M \del_- \tilde X^N~ , ~~~ 
\tilde Q^\pm_{MN}=\tilde G_{MN}\pm \tilde B_{MN}~ , ~~~ 
\tilde X^M=(\tilde X^\m,Y^i) ~ ,
\label{tSQ}
\ee
and a transformation that relates the conjugate 
momenta $P_\m$ and $\tP_\m$ to $X^\m$ and
$\tX^\m$ as well as to their first derivatives $\del_\s X^\m$ and
$\del_\s \tX^\m$. We do not include higher derivatives in $\s$ because such
derivatives do not appear in \eqn{SQ}, \eqn{tSQ}.
The most general such transformation is of the 
form (see however remarks in section 4):
\ba
P_\m & = & A_{\m\n}(X,\tX) \del_\s X^\n + C_{\m\n}(X,\tX) \del_\s \tX^\n  ~ ,
\nonumber \\
\tP_\m & = & \tilde A_{\m\n}(X,\tX) \del_\s \tX^\n
+ \tilde C_{\m\n}(X,\tX) \del_\s X^\n ~ .
\label{PPP}
\ea
Then it is a canonical transformation if one can find a generating 
functional $F$ such that
\be
P_\m = {\d F\ov \d X^\m}~ , ~~~~ \tP_\m = - {\d F\ov \d \tX^\m}~ .
\label{PFPF}
\ee
The generating functional should be of the form (we consider 
functionals with no explicit $\tau$-dependence):
\be
F= \oint d\s \left( B_\m(X,\tX) \del_\s X^\m 
+ \tilde B_\m(X,\tX) \del_\s \tX^\m \right ) ~ ,
\label{Funcc}
\ee
for some functions $B_\m$, $\tilde B_\m$ of the target space variables that 
have to be determined.
Using \eqn{PFPF}, \eqn{Funcc} we find that (we will denote
$\del_\m ={\del\ov \del X^\m}$ and $\tilde \del_\m ={\del\ov \del \tX^\m}$)
\ba 
P_\m & = & \del_{[\m} B_{\n]} \del_\s X^\n + (\del_\m \tilde B_\n -
\tilde \del_\n B_\m) \del_\s \tX^\n ~ ,
\nonumber\\
\tP_\m & = & - \tilde \del_{[\m} \tilde B_{\n]} \del_\s \tX^\n
-  (\tilde \del_\m  B_\n - \del_\n \tilde B_\m) \del_\s X^\n ~ .
\label{PMPM}
\ea
Notice that $B_\m$ and $\tilde B_\m$ may be changed by total derivatives
$\del_\m \L_2$ and $\tilde \del_\m \L_2$, respectively, with $\L_2=\L_2(X,\tX)$,
without affecting the generating functional \eqn{Funcc} and the 
transformation \eqn{PMPM}.
Next we compare \eqn{PMPM} to \eqn{PPP} and obtain the following first-order 
differential equations for $B_\m$, $\tilde B_\m$
\ba
\del_{[ \m} B_{\n ]} & = & A_{\m\n} ~ ,
\label{EQ1} \\
\tilde \del_{[ \m} \tilde B_{\n ]} & = & -\tilde A_{\m\n} ~ ,
\label{EQ2} \\
\del_\m \tilde B_\n - \tilde \del_\n B_\m 
& = & C_{\m\n} ~ ,
\label{EQ3} \\
\tilde \del_\m B_\n - \del_\n \tilde B_\m 
& = & - \tilde C_{\m\n} ~ .
\label{EQ4}
\ea
One immediately sees\footnote{Note that, 
solving \eqn{EQ1}, \eqn{EQ2} determines
$B_\m$ and $\tilde B_\m$ up to total derivatives $\del_\m \L_1(X,\tX)$ and
$\tilde \del_\m \L(X,\tX)$. We may consistently set $\L_1=0$ since 
otherwise it may
be absorbed into a redefinition of $\L$, i.e. $\L\to \L + \L_1$. 
The function $\L$ is then determined
by demanding that \eqn{EQ3} is also satisfied.}
that the matrices in \eqn{PPP} should obey
\be
A_{\m\n}=-A_{\n\m}~ , ~~~~ \tilde A_{\m\n} = - \tilde A_{\m\n}~ ,~~~~
\tilde C_{\m\n} = C_{\n\m}~ .
\label{propp}
\ee 
The latter condition makes \eqn{EQ4} identical to \eqn{EQ3}.
Further compatibility conditions 
for \eqn{EQ1}--\eqn{EQ4} may be easily derived by appropriately differentiating
them and antisymmetrizing some indices
\ba
\del_\m A_{\n\r} + \del_\r A_{\m\n} + \del_\n A_{\r\m} 
& = & 0 ~ ,
\label{compa1}  \\
\tilde \del_\m \tilde A_{\n\r} + \tilde \del_\r \tilde A_{\m\n} + 
\tilde \del_\n \tilde A_{\r\m} 
& = & 0 ~ ,
\label{compa2} \\
\del_\m \tilde A_{\n\r} + \tilde \del_\n C_{\m\r}- \tilde \del_\r C_{\m\n}
& = & 0 ~ ,
\label{compa3} \\
\tilde \del_\m A_{\n\r} + \del_\n C_{\r\m}- \del_\r C_{\n\m}
& = & 0 ~ .
\label{compa4} 
\ea
%
%
%
%
Equations \eqn{EQ3}, \eqn{EQ4} considered as second-order equations for
$B_\m$, $\tilde B_\m$ after using \eqn{EQ1}, \eqn{EQ2},
can be integrated once to give
\be 
\tilde \del_\m B_\n = - C_{\n\m} + \del_\n \tilde \G_\m ~ .
\label{us11}
\ee
and
\be 
\del_\m \tilde B_\n =  C_{\m\n} + \tilde\del_\n \G_\m ~ ,
\label{us2}
\ee
where $\tilde \G_\m$, $\G_\m$ are some unknown functions of $X^\m$, $\tX^\m$.
Equations \eqn{us11}, \eqn{us2} will be shortly used in deriving 
\eqn{AMm}, \eqn{tAMm} below, where the effect of $\tilde \G_\m$, $\G_\m$
will also be explained.

\no
Finally, one considers the Hamiltonians corresponding to \eqn{SQ} 
and \eqn{tSQ}. Namely,
\be
H=\ha \oint d\s \left( G^{MN}(P_M + B_{M\L} \del_\s X^\L) 
(P_N + B_{NP} \del_\s X^P) + G_{MN} \del_\s X^M \del_\s X^N \right) ~ ,
\label{hhh}
\ee
and a similar expression for $\tilde H$. For \eqn{PPP} to be a
canonical transformation, one must require that apart from the fact that it
can be derived from a generating functional using \eqn{PFPF},
the Hamiltonians for \eqn{SQ}, \eqn{tSQ} must be equal, i.e.
$H=\tilde H$. After some tensor manipulations we find that this 
is guaranteed, provided that the following conditions hold
\ba
&& (C\inv)^{\m\r} (Q^\pm_{\r\l} \pm A_{\r\l}) (C\inv)^{\kappa\l} 
(\tilde Q^\pm_{\kappa\n} \pm \tilde A_{\kappa\n}) = \d^\m_\n ~ ,
\nonumber \\
&& Q^\pm_{\m i} = (Q^\pm_{\m\r} \pm A_{\m\r}) 
(C\inv)^{\l\r}\tilde Q^\pm_{\l i}~ .
\label{hthc}
\ea
Hence, knowledge of the canonical transformation and of $Q^\pm_{MN}$
($\tilde Q^\pm_{MN}$) determines $\tilde Q^\pm_{MN}$ ($Q^\pm_{MN}$).

Let us return to the case of \PL. Then, as it was mentioned, the $X^\m$'s 
and the $\tX^\m$'s parametrize the groups $G$ and $\tilde G$ respectively.
It is therefore natural to use, instead of $A_{\m\n}$, 
$\tilde A_{\m\n}$ and $C_{\m\n}$, matrices with group-space indices defined
as
\be 
A_{\m\n} = A_{ab} L^a_\m L^b_\n~ , ~~~ 
 \tilde A_{\m\n} = \tilde A^{ab} \tL_{\m a} \tL_{\n b}~ , ~~~ 
 C_{\m\n} = C_a{}^b L^a_\m \tL_{\n b}~ ;
\label{aacc}
\ee
also, we identify $E^\pm_{AB}=L^A_M L^B_N Q^\pm_{MN}$ and  
$\tilde E^{\pm AB}=\tL^{MA} \tL^{NB} \tilde Q^\pm_{MN}$.
Then comparing \eqn{cantr2} and \eqn{PPP} we read off the matrices
$A_{ab}$, $\tilde A^{ab}$, $C_a{}^b$ and $\tilde C^a{}_b$ as
\ba
&& A  = (I-\tilde \Pi \Pi)\inv \tilde \Pi ~ , ~~~~ 
\tilde A =  (I- \Pi \tilde \Pi)\inv \Pi ~ ,
\nonumber \\
&& C  =  ( I - \tilde \Pi \Pi)\inv ~ ,~~~~~~
\tilde C  =  ( I - \Pi\tilde  \Pi)\inv ~ .
\label{BBCC}
\ea
It can be easily seen that \eqn{propp} and \eqn{hthc} are indeed satisfied.
Solving the differential equations \eqn{EQ1}--\eqn{EQ3} turned out to be
a much more difficult task. We found it convenient to use the 
parametrization for the group element $\tilde g=e^{i \tilde X_a \tilde T^a}$
and hence the target-space variables for the action \eqn{tsmoac}
are the $\tX_a$'s. Then it turns out that\footnote{For a general element 
$l$ of a group $D$ (with algebra generators $T_A$)
parametrized as $l=e^{iT_A X^A}$, 
one has \cite{schouten} $l\inv T_A l = (e^{-F})_A{}^B T_B$ and
$L=(1-e^{-F})/F$, where $F_A{}^B=F_{AC}{}^B X^C$.
In our case the group $D$ is the Drinfeld double, 
$l=\tilde g$ and $T_A=\{T_a,\tilde T^a\}$. Then 
using \eqn{abpi} one reads off the matrix 
$F=\pmatrix{-\tilde f^T & -f \cr
0 & \tilde f \cr}$, from which one computes $e^{-F}$ and hence
$\tilde a$, $\tilde b$, $\tilde \Pi$ and $\tL$.}
\ba 
&& \tilde a ~ = ~ e^{- \tilde f}~ =~  I - \tilde f + \ha \tilde f^2 + \dots ~ ,
\nonumber \\
&&\tilde b ~ =~  \sum_{n=0}^\infty \sum_{m=0}^n {(-1)^m\ov n!} 
(\tilde f^T)^{n-m} f \tilde f^m 
~ =~   f - \ha (f \tilde f - \tilde f^T f) + \dots ~ ,
\nonumber \\
&& \tilde \Pi ~ =~  \sum_{n=0}^\infty \sum_{m=0}^n {(-1)^{n+1}\ov (n+1) 
 m! (n-m)!} (\tilde f^T)^{n-m} f \tilde f^m 
~ =~  - f + \ha ( f \tilde f + \tilde f^T f) + \dots ~ ,
\label{paramm} \\
&& \tL~  =~ 
 (I-e^{-\tilde f})/{\tilde f} = I - \ha \tilde f + {1\ov 6} \tilde f^2
+ \dots ~ ,
\nonumber
\ea
where the matrices $f$ and $\tilde f$ have components 
\be f_{ab} = f_{ab}{}^c \tilde X_c ~ ,  ~~~~~~
\tilde f^a{}_b = \tilde f^{ac}{}_b \tX_c ~ .
\label{ftf}
\ee
Similar expressions hold for the untilded quantities in the parametrization 
of $g=e^{i X^a T_a}$.
Equations \eqn{EQ1} and \eqn{EQ2} can be solved as a power series in the 
$\tX_a$'s. In fact we may find a closed formula for $B_\m$ by
using \eqn{us11} with \eqn{BBCC} to evaluate the higher-order derivatives
in the Taylor expansion of $B_\m$ in powers of the $\tX_a$'s. 
The result is 
\be
B_\m = -L^a_\m \int_0^1 dt C_a{}^b(X,t\tX) \tX_b ~ ,
\label{AMm}
\ee
where in the matrix $C$, defined in \eqn{BBCC},
all the $\tX_a$'s contained in the expression for $\tilde \Pi$ in 
\eqn{paramm} are replaced by $t \tX_a$'s. Notice also that in accordance 
with the remaks in footnote 5 we have dropped
the contribution of the function $\tilde \G_\m$, since it is a
$\del_\m$-derivative.
A result similar to \eqn{AMm} holds for $\tilde B_\m$
\be
\tilde B_\m = \tL_{\m a} \int_0^1 dt C_b{}^a(t X,\tX) X^b + \tilde \del_\m
\L(X,\tX) ~ ,
\label{tAMm}
\ee
where we have now replaced in $C$ all the $X^a$'s 
contained in the expression for $\Pi$ 
by $t X^a$'s.\footnote{We may easily write \eqn{AMm}, \eqn{tAMm} 
in an arbitrary coordinate system by just expressing 
$\tX_a$ and $X^a$ in terms of the corresponding group elements $\tilde g$ and 
$g$ respectively. For instance, \eqn{AMm} takes the form
\be
B_\m = i L^a_\m \int_0^1 dt C_a{}^b(g,\tilde g^t) 
\langle{\ln \tilde g|T_b\rangle} ~ ,
\label{AMmcoo}
\ee
where instead of the group element $\tilde g$ in the
definition of $C$ in \eqn{BBCC} we must use its power $\tilde g^t$. 
A similar expression holds for \eqn{tAMm} as well.}
The function $\L(X,\tX)$ is related to $\G_\m$ in \eqn{us2}
and is to be determined perturbatively 
by demanding that \eqn{EQ3} is also satisfied. 
It seems helpless to exactly perform the integral in 
\eqn{AMm} (and in \eqn{tAMm}) because the $t$-dependence of $C(X,t\tX)$ 
is quite complicated.
To order $O(\tX)^2$, but exactly in the $X^\m$ variables, we find
\ba
B_\m & = & L^a_\m ( -\tX_a +\ha f_{ab} \Pi^{bc} \tX_c + \dots ) ~ ,
\nonumber \\
\tilde B_\m & = & \ha \P^{\m a} \tX_a 
+ {1\ov 6} \L^{ab\m} \tX_a \tX_b + \dots ~ ,
\label{BtB}
\ea
where $\L^{ab\m}$ is defined in \eqn{id9}. 
In the case of 
non-Abelian duality where the group $\tilde G$ 
(equivalently $G$) is Abelian, \eqn{BtB} is reduced to $B_\m = -L^a_\m \tX_a$,
$\tilde B_a =0$ (since then $\tilde f^{ab}{}_c = \Pi^{ab} =0$), which in fact
gives the exact results for the generating functional in that case 
\cite{zacloz,can3,KSsusynab}. 
Hence, the extra terms in \eqn{BtB} represent in some 
sense the deviations of \PL\ from the usual non-Abelian duality.
When the group 
$G$ is also Abelian, \eqn{BtB} is further reduced to $B_a= -\tilde X_a$,
$\tilde B_a =0$,  which 
of course corresponds to the exact generating functional for Abelian 
duality \cite{can2}.
Let us also note that since the generating functional in highly 
non-linear in $X^\m$, $\tX^\m$, it will presumably receive quantum corrections
when the canonical transformation is implemented in the full Hilbert space
along the lines of \cite{qufun}.

\subsection{The supersymmetric case}

We would like to extend the previous discussion to $N=1$ 
supersymmetric $\s$-models, with actions corresponding to \eqn{SQ}, \eqn{tSQ}
given by 
\ba
&& S_{N=1} = 
\ha \int Q^+_{MN} \del_+ X^M \del_- X^N + i Q^+_{MN} \Psi_+^M \del_-
\Psi_+^N + i Q^-_{MN} \Psi_-^M \del_+ \Psi_-^N 
\nonumber\\
&& ~~~ + i S^+_{M;\L N} \Psi^M_+ \Psi^N_+ \del_- X^\L
+ i S^-_{M;\L N} \Psi^M_- \Psi^N_- \del_+ X^\L 
+ \ha R^-_{MNP\L} \Psi^M_+ \Psi^N_+ \Psi^P_-  \Psi^\L_-  ~ ,
\label{sn1}
\ea
where $S^\pm_{M;\L N} = \ha \del_{[N} Q^\pm_{M] \L}$ and
$R^-_{MNP\L}$ is the generalized curvature tensor.\footnote{Action \eqn{sn1} 
appears slightly different 
from the standard expressions found in the literature 
(see for instance \cite{zumino}). The reason is that 
we have retained the total derivative terms 
\be
\D S = -{i\ov 2} \int \del_+(B_{MN} \Psi^M_- \Psi^N_-) - \del_-(B_{MN}
\Psi^M_+ \Psi^N_+) ~ ,
\label{totder}
\ee
which are usually dropped in passing
from the superfield to the component formalism of $N=1$ supersymmetry.
We may verify that if we subtract \eqn{totder} from \eqn{sn1} we get
the usual form of the action. This point was appreciated in the second
of refs. \cite{zasymp}.}
A similar expression holds for $\tilde S_{N=1}$ as well.
The generating functional 
is easily obtained by supersymmetrizing the bosonic one \eqn{Funcc}.
This is done by simply replacing the bosonic fields and derivatives 
by their supersymmetric counterparts. 
However, 
in order to do that we first have to take the $\tau$-derivative of \eqn{Funcc}
because this is how it relates the Lagrangians by definition,
i.e. $\cL=\tilde \cL + {d F\ov d \tau}$. 
Then one obtains 
\ba
{ d F_{N=1}\ov d \tau} & =&   -\ha \oint d\s d\th_- d\th_+ \biggl(
A_{\m\n} D_+ Z^\m D_- Z^\n 
- \tilde A_{\m\n} D_+ \tilde Z^\m D_- \tilde Z^\n 
\nonumber \\
&& ~~~~~~~~~
+ C_{\m\n} (D_+ Z^\m D_- \tilde Z^\n - D_+ \tilde Z^\n D_- Z^\m) \biggr)  ~ ,
\label{dfsusy}
\ea
where we have used \eqn{EQ1}--\eqn{EQ3}.
The (bosonic) superfield and superderivatives are given by
\ba
Z^M & = & X^M -i \th_+ \Psi_-^M + i \th_- \Psi_+^M - i \th_+ \th_- F^M  ~ ,
\nonumber\\
D_\pm & = & \mp i \del_{\th_\mp} \mp \th_\mp \del_\pm ~ ,
\label{zmd}
\ea
with $F^M= i \Om^{+M}_{N\L} \Psi_-^N \Psi^\L_+$ on shell, where 
$\Om^{+M}_{N\L}$ is the generalized connection that includes the torsion
$H_{MN\L}=\del_{[M} B_{N\L]}$ in its definition.
An expression similar to \eqn{zmd} holds for $\tilde Z^M$ as well. 
After we perform the $\th_\pm$
integration we are left with an integrand that can be written, after some
tensor manipulations, as a total derivative in 
$\tau$. From that we extract the generating functional, which assumes the form
\ba 
F_{N=1} & = & 
\oint d\s \biggl( B_\m \del_\s X^\m + \tilde B_\m \del_\s \tilde X^\m 
-{i\ov 4} A_{\m\n}(\Psi_+^\m \Psi_+^\n -\Psi_-^\m \Psi_-^\n) 
\nonumber \\
&& ~~~~~~
+{i\ov 4} \tilde A_{\m\n}(\tilde \Psi_+^\m \tilde \Psi_+^\n -\tilde \Psi_-^\m 
\tilde \Psi_-^\n) - {i\ov 2 } C_{\m\n} (\Psi_+^\m \tilde \Psi_+^\n 
-\Psi_-^\m \tilde \Psi_-^\n )\biggr)~ .
\label{FN1}
\ea
Notice that the fermionic terms in the above expression are completely
determined by the matrices entering into the bosonic transformations \eqn{PPP}.
Using \eqn{FN1}, \eqn{PFPF}
the transformation rules for $P_\m$, $\tilde P_\m$ are found to be
\ba
P_\m & = & A_{\m\n} \del_\s X^\n + C_{\m\n} \del_\s \tX^\n 
-{i\ov 4} \del_\m A_{\n\r}(\Psi_+^\n \Psi_+^\r -\Psi_-^\n \Psi_-^\r) 
\nonumber \\
&&  
+{i\ov 4} \del_\m \tilde A_{\n\r}(\tilde \Psi_+^\n \tilde \Psi_+^\r
-\tilde \Psi_-^\n \tilde \Psi_-^\r) 
- {i\ov 2 } \del_\m C_{\n\r} (\Psi_+^\n \tilde \Psi_+^\r 
-\Psi_-^\n \tilde \Psi_-^\r )  ~ ,
\label{PPn1}
\ea
with a similar expression for $\tilde P_\m$.
The transformation rules for fermions are found by using 
\be
P_{\Psi_\pm^\m} = {\d F_{N=1}\ov \d \Psi^\m_\pm}~ , ~~~~
P_{\tilde \Psi_\pm^\m} = {\d F_{N=1}\ov \d \tilde \Psi^\m_\pm} ~ ,
\label{Psipsi}
\ee
and the explicit expressions for the conjugate to the fermions
momenta, which are given by
\ba
P_{\Psi_\pm^\m} = {\d S_{N=1}\ov \d \Psi^\m_\pm} &= &
-{i\ov 2} (Q^\mp_{\m \n } \Psi_\pm^\n + Q^\mp_{\m i } \Psi_\pm^i ) ~ , 
\nonumber\\
P_{\tilde \Psi_\pm^\m} = {\d S_{N=1}\ov \d \tilde \Psi^\m_\pm} &= &
-{i\ov 2} (\tilde Q^\mp_{\m \n } \tilde \Psi_\pm^\n 
+ Q^\mp_{\m i } \Psi_\pm^i ) ~ .
\label{psimo}
\ea
The result is
\ba
\tilde \Psi_\pm^\m & = & 
\pm (C\inv)^{\m\n} \left((Q^\mp_{\n\r}\mp A_{\n\r})\Psi^\r_\pm
+Q^\mp_{\n i} \Psi_\pm^i \right) ~ ,
\nonumber \\ 
\Psi_\pm^\m & = & 
\pm (C\inv)^{\n\m} \left((\tilde Q^\mp_{\n\r}\mp 
\tilde A_{\n\r})\tilde \Psi^\r_\pm
+ \tilde Q^\mp_{\n i} \Psi_\pm^i \right) ~ .
\label{psitransf}
\ea
Consistency of the two transformations is guaranteed by \eqn{hthc}.
%
%
In the case of \PL, it is natural to use 
fermions with tangent-space indices, i.e. $\Psi^a_\pm = \Psi^\m_\pm L^a_\m$
and $\tilde \Psi_{\pm a} = \tilde \Psi^\m_\pm \tL_{\m a}$. Then using
\eqn{aacc} we may easily show that \eqn{psitransf} reduces to 
eq. (40) of \cite{PLsfe1}. 
In the case of non-Abelian duality using the fact that $\Pi^{ab}=0$,
$\tilde \Pi_{ab}= - f_{ab}{}^c \tX_c$, the generating functional \eqn{PPn1}
reduces to
\be
F^{n.a.}_{N=1} = \oint d\s \left( L^a_\s \tX_a +
{i\ov 4} f_{ab}{}^c \tX_c (\Psi^a_+ \Psi^b_+ - \Psi^a_- \Psi^b_-) -{i\ov 2}
(\Psi^a_+ \tilde \Psi_{+a} - \Psi^a_- \tilde \Psi_{-a}) \right) ~ ,
\label{n1noab}
\ee
which has been used in \cite{zasymp}.

\section{Concluding remarks and future directions}

We explicitly formulated \PL\ as a canonical transformation on the
world-sheet by proving that a certain transformation,
conjectured in \cite{PLsfe1}, between 
phase-space variables does indeed leave 
the Poisson brackets invariant.
Since the corresponding $\s$-models do not generically
have any isometries, we have provided the first example of a class of 
models that are non-isometric but still classically
canonically equivalent.\footnote{This is similar 
in spirit to being able to perform
a duality transformation in antisymmetric tensor theories in the presence of
magnetic sources or monopole condensates, where also global symmetries are
broken (see for instance \cite{quevdualglo}).
However, a canonical formulation in these cases is lacking.}
Furthermore we have presented a general discussion on generating functionals
of canonical transformations in bosonic and $N=1$ supersymmetric $\s$-models
and derived the complete set of conditions that determine them.
Applying the general formalism to the case of \PL, we derived
perturbatively, in a sense that was explained, the corresponding generating
functional.

Although not immediately obvious, the form of the canonical transformation
\eqn{PPP} is not the most general one. The reason is 
that we have excluded 
Wilson lines from appearing in the right-hand sides. 
A necessary condition for Wilson lines to be permissible 
is that they cancel each
other when the transformation is implemented into the Hamiltonian, which is 
bilinear. It is clear that in its usual formulation the generating 
functional approach to canonical transformations (cf. \eqn{PFPF}, \eqn{Funcc})
cannot accommodate these more general transformations, but the
canonical equivalence can still be shown by examining the invariance of the
Poisson brackets.
An important example of such a canonical transformation was given in 
\cite{KSsusynab} for the WZW model for a group $G$ 
and its non-Abelian dual with respect to the vector action of a subgroup
$H$. In this case the canonical mapping for the group element $g\in G$ is
\be 
g\to h_-\inv g h_- ~ , ~~~~~ h_- = P e^{-\int^{\s^-} A_-}~ ,
\label{paao}
\ee
where $P$ stands for path ordering, and
$A_-$ is identified with the on-shell value of the gauge field in 
the standard approach with Langrange multipliers and it depends on $g$ as well
as on the Lagrange multipliers themselves.
In the right-hand side of \eqn{paao} (including $A_-$)
we keep only those parameters in $g\in G$ 
and among the Lagrange multipliers that are
gauge-invariant (this is equivalent to gauge fixing $\dim(H)$ out of the total 
$\dim(G) + \dim(H)$ parameters).
Then it turns out \cite{KSsusynab} that 
the local chiral and anti-chiral 
currents of the WZW model become parafermionic-like 
objects of the same chirality, with Wilson lines attached to them.
Since the conjugate momenta in the WZW model are related 
linearly to the chiral and anti-chiral
currents, it is clear that Wilson lines will appear in the right-hand 
side of \eqn{PPP}. 
Nevertheless, the Hamiltonian is still local in the dual target-space 
variables and the affine current
Poisson bracket algebra of the original WZW model 
is indeed respected \cite{KSsusynab}.
It seems plausible that
a canonical formulation of the so-called 
quasi axial-vector duality \cite{axialvector} and of the non-Abelian duality
in PCMs with respect to the vectorial action of the group,
will utilize these more general transformations.
A detailed account of these and related issues is interesting but
beyond the scope of the present paper.

Our results are useful in discussing T-duality for open strings in 
D-manifolds. Firstly, by knowing the explicit canonical map in phase space
for \PL, we may study the mapping of the boundary conditions in a 
systematic way.
This will generalize and unify
previous work on Abelian and non-Abelian duality for open
strings \cite{DaLePo,GrHo,Tdopen} and also complement and clarify existing
work \cite{kliDbr} on D-branes and \PL.
Perhaps more importantly, we may study T-duality when the backreaction 
string loop effects
into the backgrounds fields due to the D-branes are taking into account.
These are due to a Fischler--Susskind type mechanism \cite{FiSu} which 
in general doesn't respect the isometries the original backgrounds might have,
hence rending the 
discussion within the usual context of Abelian and non-Abelian
T-duality as inappropriate.
Whether or not the backreaction-corrected backgrounds
will be related, at least in some cases, by \PL\ is an open 
problem;there is a priori no reason why they will be so. 
However, if this is the case, the corrections to the generating functional 
we are after (for the case of Abelian or non-Abelian duality 
for open strings) should take the form of the correction terms in \eqn{BtB}. 
Even if the backreaction-corrected backgrounds are not \PL-related,
we still believe that our quite 
general (proviso the remarks of the previous paragraph) treatment of
generating functionals will be useful in studying their possible canonical
equivalence and hence shed more light into D-brane physics. 
However, there could be an interesting twist in this:
since T-duality is valid at each order in the string coupling 
perturbation theory and since
the backreaction effects arise from a higher order in the
corresponding expansion than the original background, 
it is not clear whether or not T-duality is preserved by 
the combined effect or it is replaced by some other symmetry.
Seeds of such symmetry (if it exists at all) 
should be found in a non-perturbative formulation
of string theory.
It is very much worth investigating these issues further. 

For generic backgrounds the T-duality rules have to be corrected order by
order in the $\s$-model coupling constant perturbation theory.
This was first shown for the case of Abelian duality in a restricted class of 
conformal backgrounds \cite{tseabcor}. In \cite{ouggroi} a careful analysis
on the equivalence of $\s$-models, related classically by Abelian and 
non-Abelian T-duality and treated as 2-dimensional field theories in their
own right, was carried out.
The primary example used in \cite{ouggroi} was a 1-parameter family model
interpolating between the $SU(2)$ PCM and the $O(3)$ $\s$-model (2-sphere).
The upshot of their analysis was that for Abelian duality the equivalence 
holds at 1 loop, but at the 2-loop level it
requires that the classical transformation be corrected, in accordance in 
spirit with \cite{tseabcor}.
For non-Abelian duality, a similar statement holds as well. 
One should perform a similar
analysis in an example where \PL\ is non-trivially realized, i.e. it does not 
reduce to the usual Abelian or non-Abelian duality. 
The analysis of \cite{ouggroi} did not include counterterms
that would have spoiled the isometries that the classical action had. 
It will be 
interesting to investigate this issue in the spirit of \PL, or even the more
general ones discussed in section 3, by allowing certain counterterms that
will be admissible in a more general framework,
even though they will break isometries. 
Perhaps Abelian and non-Abelian duality on generic curved backgrounds are 
meaningful only classically and quantum effects force us to work
with more general models, with less isometries.
A related issue is the preservation or not of conformal invariance in
string backgrounds under \PL,
which is related to the transformation properties of the dilaton. 

Our final brief remark concerns mirror symmetry and T-duality in 
Calabi--Yau compactifications. 
Since Calabi--Yau manifolds have no isometries, 
in order to prove mirror symmetry as T-duality,
it seems imperative to understand T-duality in the absence of isometries.
Our results on \PL\ and especially our general 
formulation of T-duality in non-isometric backgrounds might be helpful towards
proving this conjecture.

\renewcommand{\theequation}{A.\arabic{equation}}

\appendix 
\section{Derivation of useful identities}
\setcounter{equation}{0}

In this appendix we prove some useful identities by using mainly
\eqn{abpi} and \eqn{conss}. All of them
have their corresponding ``duals'', obtained by just replacing tilded symbols
by untilded ones and vice versa. We start by
explicitly evaluating 
$\langle g\inv T_b g | [\tilde T^a,g\inv T_c g]\rangle = \langle \tilde
T^a| g\inv[T_c,T_b]g\rangle $. We find the identity
\be
a_b{}^c a_c{}^d f_{cd}{}^a = f_{bc}{}^d a_d{}^a ~ .
\label{id1} 
\ee
Next we evaluate
$\langle g T_b g\inv | [\tilde T^a,g \tilde T^cg\inv ] \rangle =
\langle \tilde T^a | g [ \tilde T^c,T_b] g\inv \rangle $ and contract the
result with $(a\inv)_a{}^d$ to find, after we use \eqn{id1} and relabel the
indices, the identity
\be
f_{cd}{}^{[a} \Pi^{b]d} + \tilde f^{ab}{}_c = \tilde f^{de}{}_f (a\inv)_c{}^f
a_d{}^a a_e{}^b ~ .
\label{id2}
\ee
Another identity is derived by evaluating
$\langle g\inv \tilde T^b g|[\tilde T^a,g\inv \tilde T^c g]\rangle =
\langle \tilde T^a|g\inv [\tilde T^c,\tilde T^b]g \rangle $
and contracting the result by $a_b{}^d a_c{}^e$. We find that
\be
f_{de}{}^a \Pi^{db}\Pi^{ec} - \tilde f^{a[b}{}_d \Pi^{c]d} =
\tilde f^{de}{}_f b^{fa} a_d{}^b a_e{}^c ~ .
\label{id3}
\ee
Using the definitions \eqn{abpi} we compute the derivatives
\ba
\del_\m a_a{}^b & = & - a_a{}^c L^d_\m f_{cd}{}^b ~ ,
\label{id4} \\
\del_\m b^{ab} & = & - L^d_\m ( b^{ac}  f_{cd}{}^b 
+ (a\inv)_c{}^a \tilde f^{bc}{}_d ) ~,
\label{id5} 
\ea
and then
\be
\del_\m \Pi^{ab} =  - L^c_\m (\tilde f^{ab}{}_c + f_{cd}{}^{[a}\Pi^{b]d}) ~ .
\label{id6}
\ee
For the latter there is an alternative form found by 
using \eqn{id2}. It reads
\be
\del_\m \Pi^{ab} = - L^c_\m \tilde f^{de}{}_f (a\inv)_c{}^f a_d{}^a a_e{}^b~ . 
\label{id7}
\ee
Another useful identity is derived by contracting \eqn{id2} by $\Pi^{dc}$
and then antisymmetrizing in $b,d$ and using the definition \eqn{abpi} and
the identity \eqn{id3}. The result is the cyclic identity
\be
 \L^{[abc] }\equiv  \L^{abc} + \L^{cab} + \L^{bca} = 0 ~ ,
\label{id8}
\ee
where 
\be
\L^{abc} =
\tilde f^{a[b}{}_d \Pi^{c]d} -  2 f_{de}{}^a \Pi^{db} \Pi^{ec} ~ .
\label{id9}
\ee
It turns out that \eqn{id8} is crucial in establishing the second of 
eqs. \eqn{BtB}.
Let us also mention that more general identities can be found by replacing the
group element $g\in G$ at the starting point, by a general group element
$l\in D$.

\renewcommand{\theequation}{B.\arabic{equation}}

\appendix
\section*{B~~ Manifestly duality-invariant affine algebra}  
\setcounter{equation}{0}

In this appendix we consider the manifestly \PL-invariant action of
\cite{DriDou}, with the inclusion of the extra inert 
fields $Y^i$ \cite{PLsfe1},
and derive a Poisson bracket affine algebra, which is the
analogue of \eqn{JPJP}, \eqn{tJPJP}. We will use the notation
and definitions of \cite{PLsfe1}, which we will not repeat here. 
The action in question in given by 
eq. (26) of \cite{PLsfe1}, which we rewrite in the form
\ba
S & = & I_0(l) + {1\ov 2\pi} 
\int \langle l\inv \del_\s l| R | l\inv \del_\s l \rangle 
 -2 i \langle l\inv \del_\s l | F_i\rangle  \del_\tau Y^i 
+ 2 i \langle l\inv \del_\s l |R | F_i \rangle \del_\s Y^i
\nonumber \\
&& ~~~~~~ + \langle F_i | F_j \rangle \del_\s Y^i \del_\tau Y^j 
- \langle F_i | R | F_j \rangle \del_\s Y^i \del_\s Y^j 
\label{actt} \\
 && ~~~~~~ + \left( F_{ij} + \langle F_i|R^-_a \rangle \eta^{ab} 
\langle R^+_b | F_j \rangle -\ha \langle F_i | R | F_j \rangle \right ) 
\del_+ Y^i \del_- Y^j ~ ,
\nonumber 
\ea
where we have defined 
\be 
F_i =  F_{ai}^+ \eta^{ab} R^+_b +  F_{ai}^- \eta^{ab} R^-_b ~ .
\label{fi}
\ee 
The relation of \eqn{actt} to \eqn{smoac}, \eqn{tsmoac} is a 
follows \cite{DriDou,PLsfe1}:
In the vicinity of the unit element of $D$ we may decompose \cite{alemal} the 
group element $l\in D$ as $l=\tilde h g$ or as $l=h \tilde g$, where
$h,g \in G$ and $\tilde h, \tilde g \in \tilde G$.
Using the first decomposition we may 
integrate out $\tilde h$ using the equations of motion to solve for the 
corresponding ``gauge'' field 
$A_\pm=\tilde h\inv \del_\pm \tilde h \in \tilde \cg$. The 
result is action \eqn{smoac}.
Using the second decomposition and integrating out $h$ by solving for
$\tilde A_\pm = h\inv \del_\pm h \in \cg$ we obtain instead 
action \eqn{tsmoac}.
Action \eqn{actt} is first order in $\tau$
as far as $l$ are concerned. Hence, in order to find 
the corresponding Poisson
brackets we have to apply Dirac's procedure. This is quite straightforward
with the result
\ba
\{ \langle \d l l\inv | R^\pm_a \rangle , 
\langle \d l l\inv | R^\pm_b \rangle \} & = & \mp {\pi\ov 2} \eta_{ab}
\e(\s-\s') ~ ,
\nonumber \\
\{ \langle \d l l\inv | R^+_a \rangle , 
\langle \d l l\inv | R^-_b \rangle \} & = & 0 ~ ,
\label{dirb}
\ea
where $\e(\s-\s')$ is a step function defined as $+1(-1)$ if $\s-\s'>0$
($\s-\s'<0$).
Then if we define the current
\be 
J = i  l\inv \del_\s l = J^+_a \eta^{ab} R^+_b - J^-_a \eta^{ab} R^-_b ~ ,
\label{Jpm}
\ee
we may obtain the Poisson brackets for the components $J^\pm_a$ as
\ba
\{ J^\pm_a,J^\pm_b \} & = &
- i \pi \langle [R^\pm_a,R^\pm_b] | J \rangle \d(\s-\s')
\mp \pi \eta_{ab} \d^\prime(\s-\s') ~ ,
\nonumber \\
\{ J^+_a,J^-_b \} & = &
- i \pi  \langle [R^+_a,R^-_b] | J \rangle \d(\s-\s') ~ .
\label{POPL}
\ea
The above algebra is manifestly \PL-invariant. We note that (see also 
\cite{PLsfe1}) 
the canonical transformation \eqn{cantr} and the requirement for 
2-dimensional Lorentz invariance on the world-sheet implies
\eqn{coupl} and \eqn{coupldu} as well as the equality
\be
J^\pm_a|^{l=\tilde h g}_{A_\pm} = J^\pm_a|^{l=h \tilde g}_{\tilde A_\pm} ~ ,
\label{Jsolv}
\ee
where in the left-hand side we parametrized $l\in D$ as denoted and 
inserted the classical value of $A_\pm$, and similarly in the right-hand side.
Vice versa, one can show that 
\eqn{Jsolv} implies the canonical transformation \eqn{cantr}.


\end{document}